\documentclass[letterpaper,twocolumn,10pt]{article}
\usepackage{usenix_style}
\usepackage{comment}

\usepackage{tikz}
\usetikzlibrary{shapes,arrows,positioning,calc,fit,patterns}
\usepackage{amsmath}
\usepackage{graphicx}
\usepackage{listings}
\usepackage{xcolor}
\usepackage{algorithm}
\usepackage{algpseudocode}
\usepackage{url}
\usepackage{booktabs}
\usepackage{subcaption}
\usepackage{amssymb}

\definecolor{rust_comment}{rgb}{0.5, 0.5, 0.5}
\definecolor{rust_string}{rgb}{0.2, 0.6, 0.2}
\definecolor{rust_keyword}{rgb}{0.5, 0, 0.5}
\definecolor{rust_type}{rgb}{0.2, 0.2, 0.8}

\lstdefinelanguage{Rust}{
  keywords={true, false, let, mut, if, else, while, for, in, return, match, impl, struct, fn, pub, use, mod, crate, type, trait, where, const, unsafe, loop, break},
  keywordstyle=\color{rust_keyword}\bfseries,
  ndkeywords={self, Self, Option, Some, None, Result, Ok, Err, Vec, String, usize, u64, i32, f64, HashMap, bool, FrequencyMap, Children, StaticNode, DynamicNode, Root, DataFrame},
  ndkeywordstyle=\color{rust_type}\bfseries,
  morecomment=[l]{//},      
  morecomment=[s]{/*}{*/},
  commentstyle=\color{rust_comment}\itshape,
  stringstyle=\color{rust_string},
  morestring=[b]",
  basicstyle=\ttfamily\scriptsize,
  breaklines=true,
  showstringspaces=false,
  frame=single,
  numbers=left,
  numberstyle=\tiny\color{gray},
  captionpos=b,
  escapeinside={(*@}{@*)},
}

\begin{document}

\title{\Large \bf KELP: Robust Online Log Parsing Through Evolutionary Grouping Trees}

    \author{
  {\rm Satyam Singh}\\
  StoneBuck Labs
  \and
  {\rm Sai Niranjan Ramachandran}\\
  StoneBuck Labs\thanks{Work performed at StoneBuck Labs. Author is also affiliated with TU Munich; that affiliation is unrelated to this work.}
}

\maketitle


\
\begin{abstract}
Real-time log analysis is the cornerstone of observability for modern infrastructure. However, existing online parsers are architecturally unsuited for the dynamism of production environments. Built on fundamentally static template models, they are dangerously brittle: minor schema drifts silently break parsing pipelines, leading to lost alerts and \emph{operational toil}.
We propose \textbf{KELP} (\textbf{K}elp \textbf{E}volutionary \textbf{L}og \textbf{P}arser), a high-throughput parser built on a novel data structure: the \emph{Evolutionary Grouping Tree}. Unlike heuristic approaches that rely on fixed rules, KELP treats template discovery as a continuous online clustering process. As logs arrive, the tree structure evolves, nodes split, merge, and re-evaluate roots based on changing frequency distributions. Validating this adaptability requires a dataset that models realistic production complexity, yet we identify that standard benchmarks rely on static, regex-based ground truths that fail to reflect this. To enable rigorous evaluation, we introduce a new benchmark designed to reflect the structural ambiguity of modern production systems. Our evaluation demonstrates that KELP maintains high accuracy on this rigorous dataset where traditional heuristic methods fail, without compromising throughput. Our code and dataset can be found at \url{codeberg.org/stonebucklabs/kelp}
\end{abstract}

\section{Introduction}

The effective management of multi-tenant cloud infrastructure hinges on the ability to distill vast streams of unstructured log data into structured events. For Site Reliability Engineers (SREs), a log parser is not merely a utility; it is the first line of defense in anomaly detection and compliance auditing \cite{xu2009detecting,mi2013toward,he2021survey}.\\

Historically, logs were treated as rough but mostly stable semi-structured messages. Early parsing systems assumed that delimiters, format strings, and field layouts would remain fixed for months. These assumptions were once reasonable: monolithic applications rarely changed their logging schemas; operators crafted templates by hand; and downstream analysis pipelines could rely on the persistence of those schemas.\\

Contemporary cloud deployments though look radically different. Microservices undergo continuous deployment, rolling upgrades, partial rollbacks, A/B testing, multiversion concurrency, emergency patches, traffic shifting, failover-induced behavioral discontinuities, and architecture-level refactoring. Each of these operations introduces perturbations in log formats: fields appear and disappear, identifiers swap forms, strings break, argument orders shift, exceptions wrap, container IDs change, and entire template families bifurcate without warning.\\

This leads to a widening and now structural gap. On one side are \textbf{Heuristic Parsers} \cite{he2017drain,dai2020logram,rodrigues2021clp,yu2023brain}, which offer high throughput but rely on brittle, static depth rules. A single software update that changes a log format can shatter the parsing logic, necessitating manual rule updates. On the other side are \textbf{Machine-learning Parsers }\cite{liu2022uniparser,huang2024lunar,ma2024llmparser,zhong2024logparser}, which attempt to learn semantic structures from training datasets but incur computational and resource overheads orders of magnitude above what real-time ingestion pipelines can sustain at petabyte-per-day scale. These systems typically require GPU or TPU inference, large memory footprints, and nontrivial retraining cycles, making them inoperable for environments where logs must be parsed online, under tight latency constraints, and without sacrificing throughput.\\

Reconciling these constraints is a fundamentally challenging problem, driven by factors such as

\begin{enumerate}
    \item Incidental variations, such as IDs, counters, timestamps, and routing tags, often exhibit high cardinality but carry no structural significance. In contrast, true schema evolution, manifested as new fields, structural reordering, or message bifurcation, may appear superficially similar but fundamentally alters token distributions. Classical parsers cannot reliably distinguish between these two phenomena.

    \item A production parser cannot reprocess millions of historical logs or retrain from scratch when formats shift. Every decision must be made in the critical path of ingestion under strict latency objectives.

    \item An overly aggressive parser would fragment templates, overwhelming downstream analytics with noise. A conservative parser on the other hand would merge incompatible patterns, producing corrupted feature extractions, broken anomaly detectors, faulty cardinality estimates, and misleading audit traces.

    \item Formats drift, fork, revert, and re-converge. A parser must therefore support not only structural expansions but structural contractions. Further, production identifiers (hashes, UUIDs, probe IDs, container IDs, internal request signatures) produce extremely sparse distributions. These distributions interact with structural fields in ways that defeat delimiter-based or handcrafted rules.

\end{enumerate}

Thus, a fundamentally new perspective is needed under the given scenario. We introduce \textbf{KELP}, a system designed to bridge this widening gap. While existing log parsers treat log messages as either static templates or semantic sequences to be inferred offline, KELP starts from a different premise: large-scale log parsing is fundamentally a dynamic online clustering problem.  KELP realizes this goal through the \textbf{Evolutionary Grouping Tree} (EGT), a hierarchical data structure that incrementally organizes logs by column-level redundancy patterns. Unlike traditional template trees, which assume logs adhere to fixed delimiters or manually handcrafted hierarchies, the EGT evolves structure as the workload changes. Each insertion is a lightweight, idempotent operation that updates local tree regions without re-parsing historical logs or invoking expensive retraining cycles. This enables KELP to maintain stable parse templates even as services undergo rapid iteration.\\

To distinguish variation from structure KELP introduces a real-time \textbf{Frequency Map}, which tracks token distributions at each node and column. This mechanism gives the tree the ability to “breathe’’: When token cardinalities rise in ways indicative of behavioral evolution, the tree expands to create or refine variable nodes. When distributions consolidate, the tree pulls nodes upward, merging redundant patterns and restoring structural regularity. This bidirectional evolution is central to KELP’s robustness. It reflects the system’s design philosophy: \textit{parsing logic must adapt at the same temporal scale as the software producing the logs}. By embedding lightweight statistical feedback directly into the data structure, KELP provides the adaptability of ML-driven parsers without their computational footprint. \\

Building a robust and scalable solution further required revisiting long-standing assumptions  (Section ~\ref{sec:evaluation}) on evaluation based on existing benchmarks \cite{zhu2023loghub,jiang2024large,zhang2025logbase}. Inspired by the methodology of recent systems work \cite{verma2015large,rodrigues2021clp,yang2023skypilot}, we develop a new benchmark that captures the real operational conditions under which SREs rely on parsers for anomaly detection, auditability, and postmortem analysis. 

In summary, our work makes the following contributions: 

\begin{enumerate}
    \item \textbf{Evolutionary Grouping Tree (EGT)}: A new online parsing data structure that supports idempotent writes, column-level redundancy tracking, and dynamic structural evolution.
    \item  \textbf{Robust Evolution Algorithms}: We develop the \emph{Pull}, \emph{Root Validation}, and \emph{Re-evaluation} mechanisms that govern tree restructuring, enabling KELP to adapt to schema drift in real time while preserving stability.
    \item \textbf{A New Benchmark for Parser Robustness}: We introduce an evaluation methodology that incorporates real-world software evolution, demonstrating where existing systems fail and how KELP sustains accuracy and throughput under drift.
    
\end{enumerate}

The remainder of this paper is organized as follows. Section~\ref{sec:background} provides background on log parsing and discusses related work. Section~\ref{sec:method} details the design of KELP, including the Evolutionary Grouping Tree, Frequency Map, and the algorithms for Pull, Root Validation, and Re-evaluation, with illustrative Rust code snippets. Section~\ref{sec:implementation} presents implementation details, valuation methodology and results are in Section ~\ref{sec:evaluation} and a theoretical analysis is available at Section ~\ref{sec:theory}. Finally, Section~\ref{sec:conclusion} concludes the paper and highlights future directions.

\section{Background and Related Work}
\label{sec:background}

The problem of log parsing is best understood as the inverse of log generation. In a modern production system $S$, developers instrument code with logging statements to capture runtime state. A logging statement typically consists of a constant string literal (the \emph{template}) and a set of dynamic variables (the \emph{parameters}).

\subsection{The Log Parsing Abstraction}
Formally, let $\mathcal{L} = \{l_1, l_2, \dots, l_n\}$ be a stream of log messages emitted by $S$. A parser $P$ is a function that maps a log message $l_i$ to a tuple $(T_i, V_i)$, where $T_i$ is the static template identifier and $V_i$ is the ordered list of dynamic parameters extracted from the message.

As an illustrative example, consider a system that logs connections to services:
\begin{small}
\begin{verbatim}
1. "Connected to internal service on port 8080"
2. "Connected to external service on port 443"
\end{verbatim}
\end{small}

A naïve parser might strictly interpret every distinct string token as significant, resulting in two distinct templates due to the variation between ``internal'' and ``external.'' However, a more generalized parser would induce a hierarchical template:
\begin{equation*}
    T_{gen} = \text{``Connected to } \langle \text{var} \rangle \text{ service on port } \langle \text{var} \rangle\text{''}
\end{equation*}
This hierarchical view implies that log templates are not flat strings but Directed Acyclic Graphs (DAGs) or trees of token categories. In this hierarchy, the static prefix ``Connected to'' forms the root, branching into ``internal/external'', and subsequently converging on ``service on port''.

The central challenge of online log parsing, then, is to discover this latent structure $T$ given only the stream $\mathcal{L}$, without access to the source code, and to do so in a single pass with minimal latency.

\begin{figure*}[h] 
    \centering
    
    \includegraphics[width=\textwidth]{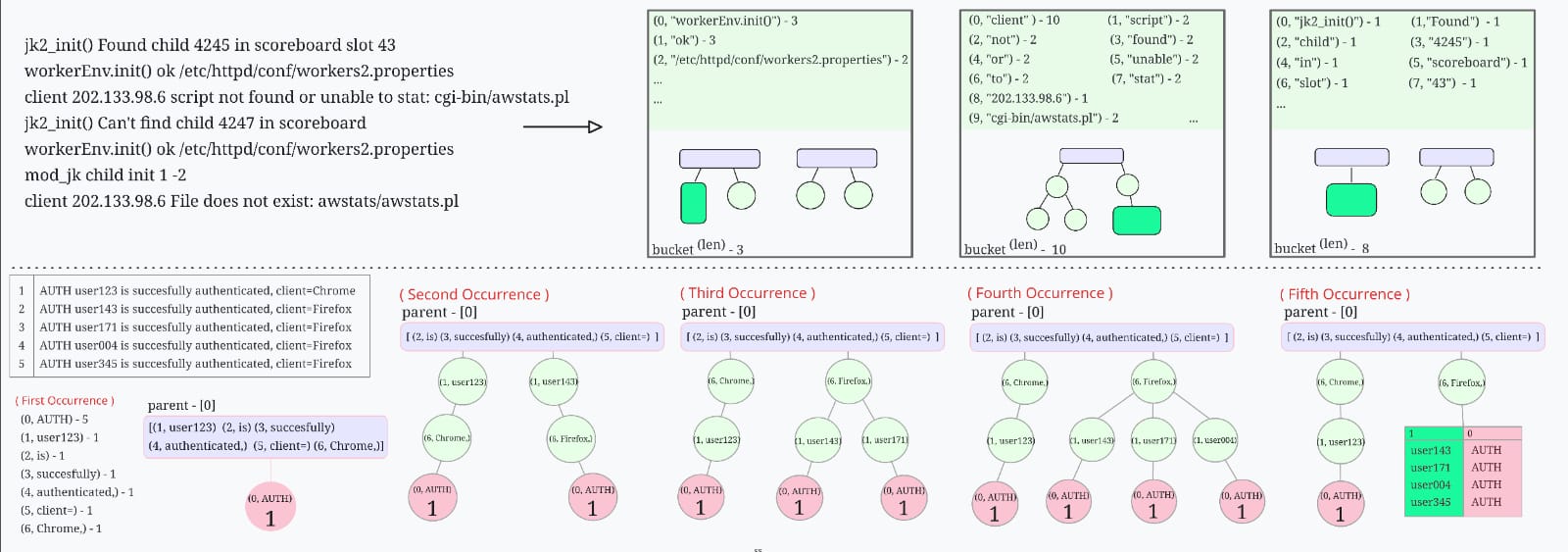}
    
    \caption{\textbf{KELP System Architecture.} Logs flow through length segregation into specific buckets. Each bucket maintains a Frequency Map and an Evolutionary Grouping Tree (EGT). The Re-evaluation Engine periodically optimizes the tree structure.}
    \label{fig:architecture}
\end{figure*}

\subsection{Related Work}
The landscape of log parsing is dominated by offline algorithms and static heuristics.

\textbf{Offline and Static Heuristics.} Early approaches like SLCT~\cite{vaarandi2003data} and LFA~\cite{nagappan2010abstracting} rely on offline, multi-pass frequent itemset mining, suffering from $O(N^2)$ complexity. While more recent adaptations like Drain~\cite{he2017drain} and Spell~\cite{du2016spell} improve throughput, they rely on rigid assumptions: Drain uses fixed-depth trees that fracture under variable-length prefixes, while Spell's LCS computation creates latency spikes in high-diversity streams. Crucially, their parsing logic remains static; they cannot structurally adapt to shifting formats without manual intervention.

\textbf{Offline-Training Heavy ML.} Deep Learning models (NuLog~\cite{nedelkoski2020self}) and LLM-based parsers~\cite{ma2024llmparser} offer semantic robustness but are operationally impractical. They necessitate heavy offline training phases and expensive inference overheads (often requiring GPUs) that are incompatible with the strict latency budgets of real-time ingestion pipelines.

A critical failure mode shared by both categories is \emph{Concept Drift}. When software evolves (e.g., an error message appends a new field), static heuristics misclassify the log as a new template, while ML models suffer distribution shift requiring retraining. Existing systems effectively treat parsing as static classification. KELP addresses this by treating parsing as \emph{online dynamic clustering}, expanding and contracting the template tree in real-time response to stream entropy.

\section{System Design}
\label{sec:method}

\subsection{System Overview}
The high-level architecture of KELP is illustrated in Figure~\ref{fig:architecture}. The pipeline operates in three stages:
\begin{enumerate}
    \item \textbf{Segregation:} Incoming logs are tokenized and bucketed by token count. This optimization, shared by systems like Drain~\cite{he2017drain}, assumes that logs generated by the same logging statement typically maintain a constant length.
    \item \textbf{Evolution:} Within each bucket, logs are processed by an EGT. KELP updates a local frequency map, validates the tree root, and pushes data into leaf nodes.
    \item \textbf{Restructuring:} Periodically (or batch-wise), KELP triggers a \emph{Re-evaluation} pass to correct structural drifts, merging fragmented branches or extracting new static columns.
\end{enumerate}

\subsection{Data Representation: The Frequency Map}
\label{subsec:freqmap}
A core insight of KELP is that a logging statement consistently places constant tokens in specific column positions. To capture this, we maintain a \textbf{Frequency Map} per bucket. This map tracks the cardinality of every token at every column index.

Consider the following stream:
\begin{small}
\begin{verbatim}
1. "Connected to client Sid on port 8080"
2. "Connected to client Luke on port 8000"
\end{verbatim}
\end{small}

The Frequency Map records the state shown in Table~\ref{tab:freq_map}. This map serves as the "ground truth" for the tree: columns with low cardinality and high frequency (e.g., "Connected", "port") are candidates for static nodes, while high-cardinality columns (e.g., "Sid", "Luke") are identified as dynamic variables.

\begin{table}[h]
\centering
\small
\begin{tabular}{|l|l|c|}
\hline
\textbf{Col} & \textbf{Token} & \textbf{Count} \\ \hline
0 & "Connected" & 2 \\ \hline
2 & "client" & 2 \\ \hline
3 & "Sid" & 1 \\
3 & "Luke" & 1 \\ \hline
5 & "port" & 2 \\ \hline
\end{tabular}
\caption{\textbf{Frequency Map State.} Tracks token distributions to distinguish static template parts from dynamic variables.}
\label{tab:freq_map}
\end{table}

\subsection{The Evolutionary Grouping Tree (EGT)}
The EGT is a hierarchical structure that organizes logs by common subsequences. Unlike fixed-depth trees, the EGT has variable depth and node types:

\begin{enumerate}
    \item \textbf{Root Node:} Represents the longest subsequence of columns that are currently considered static (based on the Frequency Map).
    \item \textbf{Static Node:} An internal node representing a specific token value at a specific column (e.g., $Col_2 = \text{"client"}$).
    \item \textbf{Dynamic Node:} A leaf node acting as a container for raw log rows. These nodes represent ambiguous or variable data that has not yet been "pulled" into a static structure.
\end{enumerate}

The tree guarantees that any path from Root to a Leaf represents a specific log pattern. However, because the data is streaming, a node that starts as a Dynamic container may later evolve into a set of Static branches.

\subsection{Primitive Operation: Pulling}
\label{subsec:pulling}
The fundamental operation for tree evolution is \textbf{Pulling}. This is analogous to a split operation in decision trees. When a Dynamic Node accumulates data, KELP inspects specific columns. If a column exhibits low cardinality, it is "pulled" up, converting the Dynamic Node into branching Static Nodes.

\begin{figure}[h]
\centering
\begin{tikzpicture}[
  node distance=0.6cm, 
  box/.style={draw, rectangle, minimum width=1.5cm, minimum height=0.6cm, align=center, font=\scriptsize},
  arrow/.style={->, >=stealth}
]
\node[box, fill=blue!10] (dyn) {Dynamic Node\\ \{Col A, Col B\}};
\node[below=0.3cm of dyn] (arrow) {Pull(Col A)};
\draw[arrow] (dyn) -- (arrow);
\node[box, fill=green!10, below left=0.8cm and -0.5cm of arrow] (stat1) {Static: "Val 1"};
\node[box, fill=green!10, below right=0.8cm and -0.5cm of arrow] (stat2) {Static: "Val 2"};
\node[box, fill=blue!10, below=0.3cm of stat1] (child1) {Dynamic\\ \{Col B\}};
\node[box, fill=blue!10, below=0.3cm of stat2] (child2) {Dynamic\\ \{Col B\}};

\draw[arrow] (arrow) -- (stat1);
\draw[arrow] (arrow) -- (stat2);
\draw[arrow] (stat1) -- (child1);
\draw[arrow] (stat2) -- (child2);
\end{tikzpicture}
\caption{\textbf{The Pull Operation.} Extracting a column from a Dynamic Node creates branching Static Nodes, refining the template structure.}
\label{fig:pulling}
\end{figure}
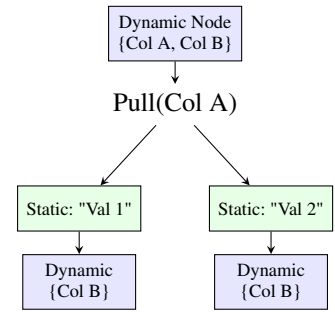

Algorithmically, \texttt{Pull} extracts a column from the raw dataframe, groups row indices by unique values, and creates new Static Nodes for each unique token. This operation is recursive: if a column is pulled from deep within a hierarchy, the operation bubbles up, restructuring the tree to surface the discriminating column.

\newpage

\begin{lstlisting}[language=Rust, caption={Recursive Pull Logic}, label={lst:pull_code}]
fn pull(&mut self, col: usize) -> Vec<StaticNode> {
    if self.col() == col { return vec![self]; }
    
    // Recursively pull from children
    self.children.pull(col).into_iter().map(|mut node| {
        // Prepend current node to pulled node's children
        node.child.prepend(self.col, self.value);
        node
    }).collect()
}
\end{lstlisting}

\subsection{Streaming Ingestion and Root Validation}
As logs arrive, KELP blindly pushes them into the tree. However, new data updates the Frequency Map, which may invalidate the current Root definition. For example, a token initially thought to be static (e.g., "User: Admin") might degrade into a variable if "User: Guest" appears later.

KELP implements \textbf{Root Validation} on every batch. It calculates a dynamic frequency threshold to separate signal from noise. We determine this threshold by analyzing the decay of the top-3 token frequencies using a natural log heuristic:
\begin{equation}
    Threshold = \left\lfloor \frac{e^{slope(top\_3)} + 1}{2} \right\rfloor
\end{equation}

If the current root columns fall below this threshold relative to the global maximum, they are demoted. The \texttt{validate\_root} function (Listing~\ref{lst:root_val}) ensures the tree's backbone always represents the most stable features of the log stream.

\begin{lstlisting}[language=Rust, caption={Dynamic Root Validation}, label={lst:root_val}]
fn validate_root(&mut self, map: &FrequencyMap) {
    // Group columns by frequency
    let chains = group_by_frequency(self.values, map);
    let threshold = calculate_threshold(chains);
    
    // Find longest sequence above threshold
    let (best_freq, best_cols) = find_best_root(chains, threshold);
    
    if best_cols != self.current_cols {
        self.re_eval = true; // Mark for restructuring
        self.demote_cols(best_cols);
    }
}
\end{lstlisting}

\subsection{Tree Restructuring: Re-evaluation}
\label{subsec:reeval}
The core of KELP's robustness is the \textbf{Re-evaluation} mechanism. Pushing data into the tree is an optimistic operation; over time, the tree structure may become suboptimal. For example, a "Static" node might fracture into thousands of children, indicating it is actually a high-cardinality variable.

The \texttt{re\_eval} process (Listing~\ref{lst:reeval}) recursively traverses the tree to restore invariants:
\begin{itemize}
    \item \textbf{Collapse (Generalization):} If a set of branches exceeds a cardinality threshold (high entropy), they are merged back into a single Dynamic Node. This effectively "learns" a wildcard.
    \item \textbf{Pull (Specialization):} If a Dynamic Node contains a column with low cardinality (low entropy), it is "pulled" to create specific Static branches.
\end{itemize}

\begin{lstlisting}[language=Rust, caption={Tree Re-evaluation}, label={lst:reeval}]
fn reeval(branches, threshold) -> Children {
  // Base Case: Merge leaves
  if branches.all_leaf() { return merge(branches); }

  // Find column with min cardinality
  let (col, words) = find_min_cardinality(branches);

  // High Entropy -> Convert to Dynamic Variable
  if words.len() > threshold {
    return Children::Dynamic(
        branches.into_iter()
            .map(|b| b.into_dynamic())
            .reduce(|a, b| a.append(b))
    );
  }

  // Low Entropy -> Pull column to form Static branches
  let split = branches.pull(col);
  Children::Static(split.map(|head, kids| {
      head.with_child(reeval(kids, threshold))
  }))
}
\end{lstlisting}

\subsection{Memory Management: Trimming}
To support infinite streams, KELP cannot store historical raw data indefinitely. We implement a "forgetful" strategy. Once a Dynamic Node reaches a capacity limit (e.g., $k$ lines), or based on a time-window, the raw data dataframe is discarded or "rolled over." The structural knowledge (the tree nodes) and statistical knowledge (the Frequency Map) are retained, ensuring the parser remains accurate without unbounded memory growth.

\section{Implementation}
\label{sec:implementation}

We implemented KELP in Rust ($\approx$3,500 LOC). The implementation strategy prioritizes memory compactness and instruction cache locality, avoiding the pointer indirection overhead typical of object-oriented parsers. The system architecture rests on three tightly coupled mechanisms designed to sustain high-throughput ingestion.

\subsection{Columnar Compression via RleVec}
A fundamental characteristic of production logs is high sequential redundancy. In a stream of 10,000 requests, the "Status Code" column might contain the integer \texttt{200} for 99\% of entries. Naïve implementations storing logs as `Vec<Vec<String>>` incur massive overhead: a 4-byte string requires a 24-byte vector header plus heap allocation, resulting in fragmentation and poor cache locality.

To mitigate this, KELP implements a custom \texttt{RleVec} (Run-Length Encoded Vector) as the backing store for `DynamicNode` leaves.
\begin{lstlisting}[language=Rust]
struct RleRun<T> { len: usize, val: T }
struct RleVec<T> { inner: Vec<RleRun<T>> }
\end{lstlisting}
The `RleVec` provides a dense, columnar storage layout. When pushing a new log token, the system checks the tail of the vector. If the token matches the previous entry (a pointer comparison), we simply increment a `usize` counter. This reduces the write complexity for steady-state logs to a single arithmetic instruction. 

Furthermore, `RleVec` supports efficient "splitting." When the Tree Re-evaluation algorithm (§\ref{sec:evaluation}) determines a column must be pulled, the `RleVec` can be sliced and reorganized without deep copying the underlying token data, only manipulating the lightweight `RleRun` headers.

\subsection{Zero-Copy Interning with Bi-Directional Maps}
String comparison is the dominant cost in log parsing. To eliminate this from the critical path, KELP implements a global interning layer using a specialized bidirectional map backed by a `Slab` allocator.

\begin{lstlisting}[language=Rust]
pub struct GlobalFreqMap {
    ids: Slab<()>, // O(1) slot allocator
    pub map: BiMap<String, TokenValueRef>,
    pub col_freq: Vec<HashMap<TokenValue, usize>>,
}
\end{lstlisting}

\textbf{Ingestion Path:} When a log line arrives, strings are hashed and looked up in the `BiMap`. If present, we return a `TokenValueRef` (a lightweight wrapper around `Arc<usize>`). If absent, the string is allocated once, inserted into a free slot in the `Slab`, and mapped. 

\textbf{Parsing Path:} Once tokenized, all internal tree operations, i.e, branch traversal, equality checks, and frequency counting operate exclusively on 8-byte integers (`usize`). This fits the "hot path" data structures entirely within L1/L2 cache, providing orders-of-magnitude faster comparisons than `strcmp`.

\textbf{Reconstruction Path:} For template generation, the `BiMap` allows $O(1)$ reverse lookup to reconstruct the original string from the integer ID.

\subsection{Polymorphic Node Layout}
Traditional parsers often use class hierarchies (v-tables) to represent tree nodes, incurring dynamic dispatch overhead on every traversal step. KELP leverages Rust's `enum` to create a memory-safe tagged union:

\begin{lstlisting}[language=Rust]
pub enum ChildEither {
    Count(usize),           
    Static(Vec<StaticNode>), 
    Dynamic(DynamicNode),   
}
\end{lstlisting}

This structure encodes the lifecycle of a log cluster directly into the type system:
\begin{enumerate}
    \item \textbf{Count (Zero-Overhead):} When a template is perfectly matched (no variance), the leaf node collapses into a single `usize` counter. This is the most compact representation possible.
    \item \textbf{Dynamic (Accumulation):} When new, ambiguous data arrives, the node transitions to `Dynamic`, using `RleVec` to buffer raw data for statistical analysis.
    \item \textbf{Static (Branching):} Once variance is confirmed via the `Pull` algorithm, the node transitions to `Static`, creating explicit branches.
\end{enumerate}
This design ensures that stable regions of the tree incur negligible memory overhead, while complex regions pay only for the necessary structure.

\subsection{Memory Lifecycle and Garbage Collection}
In a streaming system, the set of active vocabulary (IPs, Request IDs) is unbounded. An interner that only inserts would eventually exhaust memory. KELP implements a reference-counting mechanism tied to the `FrequencyMap`.

When the tree performs a `Trim` operation (discarding old logs to maintain a sliding window), it decrements the reference counts of the associated tokens. When a token's frequency across all columns drops to zero, KELP reclaims the slot in the `Slab` and removes the entry from the `BiMap`. This ensures that the memory footprint of the parser is proportional to the \emph{active} working set of templates, rather than the total history of the log stream.

\section{Evaluation}
\label{sec:evaluation}

We evaluate KELP against three baselines: \textbf{Drain}~\cite{he2017drain}, \textbf{Logram}~\cite{dai2020logram}, and \textbf{LogMine}~\cite{hamooni2016logmine}. We restrict our comparison to these heuristic parsers, as Deep Learning alternatives incur prohibitive inference overheads and offline retraining cycles that are incompatible with the strict latency constraints of real-time log ingestion. 

\subsection{Benchmarking Methodology: The Zero-Bias Protocol}
A primary contribution of this work is the identification of a structural methodology gap in existing log parsing research. We argue that standard benchmarks, specifically Loghub~\cite{zhu2023loghub,jiang2024large}, suffer from \emph{Ground Truth Leakage}, rendering them unsuitable for evaluating the robustness of online parsers in streaming production environments.

\textbf{Ground Truth Leakage.}
Loghub datasets are annotated using semi-automated processes that rely heavily on domain-specific regular expressions (e.g., specifically matching IP addresses, UUIDs, or date formats). Consequently, the "ground truth" templates are implicitly coupled with specific tokenization rules.
Evaluating a parser on Loghub often becomes a test of \emph{regex coverage} rather than \emph{structural inference}. Parsers that implement similar pre-processing rules to the annotators score artificially high, while those attempting to learn structure from raw distribution are penalized for "missed" variables that are statistically indistinguishable from static text in small samples.

To illustrate the severity of ground truth leakage, consider the following log entry from the Linux dataset in Loghub:

\begin{small}
\begin{verbatim}
Log: "authentication failure; logname= uid=0 euid=0 
      tty=NODEVssh ruser= rhost=207.243.167.114 user=root"

GT:  "authentication failure; logname= uid=<*> euid=<*> 
      tty=NODEVssh ruser= rhost=<*> user=<*>"
\end{verbatim}
\end{small}

The ground truth explicitly marks \texttt{uid=0} as a dynamic variable (\texttt{<*>}). However, within the specific scope of the dataset, this template often appears with \textbf{zero variance}, \texttt{uid} is always \texttt{0}. 

For a statistical or distributional parser, the token \texttt{0} has an entropy of zero; it is statistically indistinguishable from static keywords like \texttt{"failure"}. Consequently, any parser relying purely on data evidence \emph{must} classify it as static. The Loghub ground truth is derived not from the dataset's distribution, but from external semantic priors (i.e., the human knowledge that "UID" implies variation) or pre-baked regex rules. Evaluating distributional parsers against semantic ground truths penalizes them for lacking "oracle" knowledge that does not exist in the raw stream.

\textbf{The Solution: Synthetic Injection.}
To isolate the parsing algorithm's performance from its pre-processing heuristics, we constructed a "Zero-Knowledge" benchmark. Our goal was to create a dataset where the distinction between static templates and dynamic variables is defined solely by \emph{token cardinality}, not by token format.

We constructed the benchmark via a three-stage pipeline:
\begin{enumerate}
    \item \textbf{Template Extraction:} We extracted 165-180 unique, real-world event templates from the Apache, BSL, and Linux datasets within Loghub. This ensures the \emph{structural complexity} (message length, word position) remains representative of production systems.
    \item \textbf{Variable Erasure:} We stripped the original dynamic variables from these templates, replacing them with generic placeholders.
    \item \textbf{High-Entropy Injection:} We generated synthetic log streams by injecting high-cardinality random strings into the variable slots.
\end{enumerate}

\textbf{The Zero-Bias Constraint.}
Crucially, our evaluation protocol forbids the use of domain-specific pre-processing (e.g., "replace all numbers with \texttt{*}"). Parsers must ingest the raw, high-entropy logs. This forces the system to rely exclusively on distributional signals i.e, frequency, cardinality, and position to discover the latent structure. This mimics the \emph{cold start} \cite{qiao2024caching} problem in multi-tenant SaaS environments where SREs cannot manually craft regexes for every new tenant's log format.

We generated three datasets (Synthetic-1, -2, -3) with increasing variable complexity to stress-test the parsers' ability to distinguish signal (templates) from noise (high-cardinality variables).

\subsection{Results}

\textbf{Dataset 1: Synthetic-1 (Baseline).}
Table~\ref{tab:syn1} presents the performance on the baseline dataset. KELP matches Drain's execution time (0.22s) but significantly outperforms it in accuracy. Drain identifies 1,091 templates against a ground truth of 180, leading to a low F1 Group Accuracy (FGA) of 0.223. This confirms that without regex assistance, Drain's fixed-depth heuristic collapses under high-cardinality data. KELP maintains structural coherence with an FGA of 0.817.

\begin{table}[h]
\centering
\small
\resizebox{\columnwidth}{!}{%
\begin{tabular}{l|c|c|c|c|c|c}
\toprule
\textbf{Parser} & \textbf{Time (s)} & \textbf{ID Templates} & \textbf{GT} & \textbf{GA} & \textbf{PA} & \textbf{FGA} \\ 
\midrule
Drain    & 0.40 & 1091 & 180 & 0.790 & 0.766 & 0.223 \\
LogMine  & 37.11 & 232  & 180 & 0.863 & 0.870 & 0.752 \\
Logram   & 0.65 & 544  & 180 & 0.778 & 0.593 & 0.384 \\
\textbf{KELP}     & \textbf{0.22} & \textbf{207}  & 180 & \textbf{0.878} & \textbf{0.912} & \textbf{0.817} \\
\bottomrule
\end{tabular}
}
\caption{\textbf{Synthetic-1 Results.} KELP matches Drain's speed while avoiding template explosion.}
\label{tab:syn1}
\end{table}

\textbf{Dataset 2: Synthetic-2 (Intermediate).}
In Table~\ref{tab:syn2}, KELP achieves near-perfect Parse Accuracy (0.956). LogMine offers competitive accuracy but incurs a $160\times$ latency penalty (35.27s vs 0.22s), rendering it unsuitable for streaming. Drain continues to over-partition (1057 templates). This demonstrates KELP's ability to maintain precision even as variable distributions become more complex.

\begin{table}[h]
\centering
\small
\resizebox{\columnwidth}{!}{%
\begin{tabular}{l|c|c|c|c|c|c}
\toprule
\textbf{Parser} & \textbf{Time (s)} & \textbf{ID Templates} & \textbf{GT} & \textbf{GA} & \textbf{PA} & \textbf{FGA} \\ 
\midrule
Drain    & 0.40 & 1057 & 165 & 0.812 & 0.803 & 0.219 \\
LogMine  & 35.27 & 223  & 165 & 0.888 & 0.911 & 0.747 \\
Logram   & 0.62 & 511  & 165 & 0.819 & 0.641 & 0.405 \\
\textbf{KELP}     & \textbf{0.22} & \textbf{182}  & 165 & \textbf{0.905} & \textbf{0.956} & \textbf{0.853} \\
\bottomrule
\end{tabular}
}
\caption{\textbf{Synthetic-2 Results.} KELP achieves 0.956 Parse Accuracy with minimal overhead.}
\label{tab:syn2}
\end{table}

\textbf{Dataset 3: Synthetic-3 (High Entropy).}
Table~\ref{tab:syn3} represents the highest difficulty tier. Here, LogMine suffers a catastrophic failure, identifying only 1 template (0.0 accuracy), likely due to its clustering algorithm failing to converge on the high-entropy noise. Drain remains consistent but fragmented (753 templates). KELP proves robust, identifying 181 templates (GT: 165) with high accuracy (0.915 PA), validating the stability of the EGT's re-evaluation mechanism under maximum entropy.

\begin{table}[h]
\centering
\small
\resizebox{\columnwidth}{!}{%
\begin{tabular}{l|c|c|c|c|c|c}
\toprule
\textbf{Parser} & \textbf{Time (s)} & \textbf{ID Templates} & \textbf{GT} & \textbf{GA} & \textbf{PA} & \textbf{FGA} \\ 
\midrule
Drain    & 0.39 & 753 & 165 & 0.816 & 0.806 & 0.294 \\
LogMine  & 0.07 & 1   & 165 & 0.000 & 0.000 & 0.000 \\
Logram   & 0.58 & 577 & 165 & 0.745 & 0.608 & 0.340 \\
\textbf{KELP}     & \textbf{0.21} & \textbf{181} & 165 & \textbf{0.899} & \textbf{0.915} & \textbf{0.867} \\
\bottomrule
\end{tabular}
}
\caption{\textbf{Synthetic-3 Results.} LogMine fails; KELP maintains consistency.}
\label{tab:syn3}
\end{table}

Across all three datasets, KELP provides the only viable balance of throughput and accuracy for production streaming. It processes logs $\approx 2\times$ faster than Drain while maintaining template counts within 10-15\% of ground truth. This confirms that KELP solves the "Template Explosion" problem that plagues fixed-depth heuristics without incurring the latency costs of more complex clustering methods.

\section{Theoretical Analysis}
\label{sec:theory}

This section assumes some familiarity with martingale theory \cite{williams1991probability} to formalize convergence bounds; readers primarily interested in system implementation and results may skip to Section ~\ref{sec:conclusion} without loss of continuity. To rigorously understand the latency characteristics of KELP versus nested parsers (like Drain), we model the template discovery process as a \textbf{stochastic stopping time problem}. We employ martingale theory to derive the expected number of log lines $N$ required to fully identify a template with $m$ dynamic variables.

\subsection{Problem Formulation}
Let a log template $T$ contain $m$ dynamic tokens. For a parser to identify a dynamic token $i$, it must observe at least $\tau$ distinct values (the branching threshold).
Let $X_t$ be the number of distinct values observed for a token position after reading $t$ lines. The parsing process \emph{stops} (converges) when $X_t \ge \tau$ for all $m$ dynamic positions.

We contrast two architectural models:
\begin{enumerate}
    \item \textbf{Nested Partitioning (e.g., Drain):} The parser splits the dataset based on the value of token $i$ before analyzing token $i+1$. The probability space for token $i+1$ is conditional on the subset defined by tokens $1 \dots i$.
    \item \textbf{Parallel Identification (KELP):} The parser tracks token distributions globally (via the Frequency Map) or independently per column. The identification of token $i$ does not shrink the sample size for token $j$.
\end{enumerate}

\subsection{Martingale Analysis of Nested Partitioning}
Consider a template with $m$ dynamic fields, where each field $i$ takes one of $k$ possible values with uniform probability $p = 1/k$.
In a nested parser, identifying the $d$-th dynamic token requires analyzing a subset of logs $S_d \subset \mathcal{L}$. The probability that a random log line falls into the correct path to reach depth $d$ is:
\begin{equation}
    P(\text{reach depth } d) = \prod_{i=1}^{d} p_i = \left(\frac{1}{k}\right)^d
\end{equation}
Let $Y_n^{(d)}$ be the count of lines reaching depth $d$ after $n$ total lines. This forms a sub-martingale where $\mathbb{E}[Y_{n+1}^{(d)} | Y_n^{(d)}] = Y_n^{(d)} + (1/k)^d$.
By the \textbf{Optional Stopping Theorem} \cite{williams1991probability}, the expected number of lines $\mathbb{E}[N]$ required to accumulate $\tau$ samples at depth $m$ is:
\begin{equation}
    \mathbb{E}[N_{\text{nested}}] \approx \tau \cdot \left(\frac{1}{P(\text{reach depth } m)}\right) = \tau \cdot k^m
\end{equation}
This reveals an \textbf{exponential complexity} with respect to template depth. As dynamic complexity ($m$) increases, the data requirement explodes, explaining the \emph{template explosion} seen in Drain on high-entropy datasets (Synthetic-3).

\subsection{Martingale Analysis of Parallel Identification}
In KELP, the Frequency Map tracks column statistics independently. The identification of column $i$ as dynamic depends only on the global marginal probability $p_i = 1/k$, not on the conditional path.
The process completes when all $m$ independent random walks cross the threshold $\tau$.
Let $N_i$ be the time to identify token $i$. Since distributions are independent:
\begin{equation}
    \mathbb{E}[N_{\text{parallel}}] = \mathbb{E}[\max(N_1, \dots, N_m)]
\end{equation}
For uniform probabilities, $\mathbb{E}[N_i] = \tau \cdot k$. Modeling the identification time $N_i$ for each token as an i.i.d. exponential variable with mean $\mu = \tau k$, standard results in order statistics \cite{david2004order} establish that the expected maximum scales with the harmonic number
\begin{equation}
    \mathbb{E}[N_{\text{parallel}}] \approx \tau \cdot k \cdot H_m \approx \tau \cdot k \cdot \ln(m)
\end{equation}
where $H_m$ is the $m$-th harmonic number.

\begin{figure}[h]
    \centering
    \includegraphics[width=0.9\linewidth]{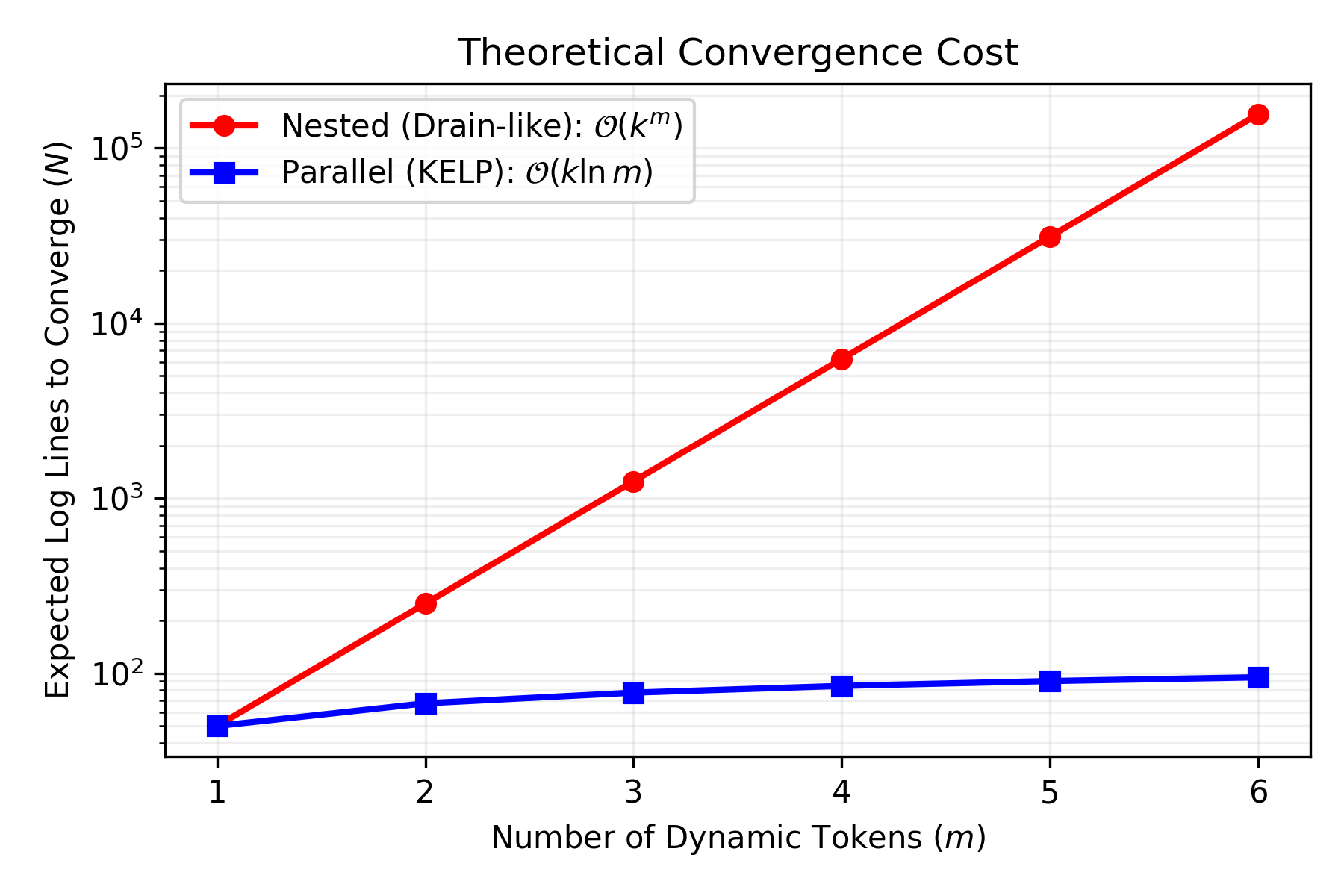}
    \caption{\textbf{Convergence Complexity.} Theoretical comparison of log lines required to identify a template. Nested approaches (red) face exponential data requirements as template complexity ($m$) grows, while KELP's parallel approach (blue) remains near-linear.}
    \label{fig:convergence}
\end{figure}

This analysis proves a fundamental advantage of KELP's design. While nested parsers suffer from $\mathcal{O}(k^m)$ data hunger, requiring exponentially more logs to converge on deep templates, KELP's parallel evolution scales as $\mathcal{O}(k \ln m)$. Figure~\ref{fig:convergence} visualizes this divergence, confirming why KELP maintains stability on the Zero-Bias Benchmark where nested heuristics fail.

\section{Conclusion and Future Work}
\label{sec:conclusion}

This paper argues that the brittleness of modern log parsing infrastructure stems from a fundamental mismatch between static parsing algorithms and dynamic production environments. We introduced \textbf{KELP}, a system that re-frames parsing as an online evolutionary process. By combining the \textbf{Evolutionary Grouping Tree} with real-time frequency analysis, KELP enables templates to expand and contract organically as software behavior shifts. Our contributions are threefold: we exposed the ground truth leakage in standard benchmarks like Loghub, necessitating a rigorous \textbf{Zero-Bias Benchmark}; we demonstrated empirically that KELP matches heuristic throughput while preventing template explosion; and we provided a theoretical martingale analysis proving that KELP's convergence complexity scales logarithmically ($\ln m$) with template depth, whereas nested approaches suffer exponential degradation ($k^m$).

While these results establish KELP as a robust foundation for autonomous observability, our work illuminates several avenues for future refinement. First, our reliance on length-based segregation can fragment templates containing multi-word variables across disparate buckets, a limitation we plan to address via a cross-bucket reconciliation layer. Second, the greedy evolutionary logic may occasionally settle on local optima such as promoting low-variance variables to ``False Roots'' which we aim to mitigate by analyzing static node chains to retrospectively correct structural misalignments. Finally, we envision extending KELP's single-node efficiency to a distributed architecture by sharding the Frequency Map and EGTs, enabling horizontal scaling for hyperscale environments.

\bibliographystyle{plain}
\bibliography{references}

\end{document}